\documentstyle[12pt]{article}
\begin{document}
\begin{titlepage}
\title{Canonical Structure of the Non-Linear $\sigma$-Model in a
Polynomial Formulation}
\author{C. D. Fosco$^1$ and T. Matsuyama$^2$
\\ \\ University of Oxford\\
Department of Physics, Theoretical Physics\\
1 Keble Road, Oxford OX1 3NP, UK }
\vspace{1cm}
\baselineskip=21.5pt
\begin{abstract}
We study the canonical structure of the $SU(N)$ non-linear $\sigma$-model
in a polynomial, first-order representation. The fundamental variables
in this description are a non-Abelian vector field $L_{\mu}$ and a non-Abelian
antisymmetric tensor field $\theta_{\mu \nu}$, which constrains $L_{\mu}$ to
be a `pure gauge' ($F_{\mu \nu}(L) = 0$) field.
The second-class constraints that appear as a consequence of the first-order
nature of the Lagrangian are solved, and the
reduced phase-space variables explicitly found. We also
treat the first-class constraints due to
the gauge-invariance under transformations of the antisymmetric tensor
field, constructing the corresponding most general gauge-invariant functionals,
which are used to describe the dynamics of the physical
degrees of freedom.
We present these results in $1+1$, $2+1$ and $3+1$ dimensions,
mentioning some properties of the $d+1$-dimensional case.
We show that there is a kind of duality between this description of the
non-linear $\sigma$-model and the massless Yang-Mills theory. This
duality is further extended to more general first-class systems.

\footnotetext[1]{Address from 15th January 1995: High Energy Group, 
International Centre for Theoretical Physics, P.O. Box 586, 34100 
Trieste, ITALY}
\footnotetext[2]{Permanent Address: Department of Physics, Nara University of
Education, Takabatake-cho, Nara 630, JAPAN}
\end{abstract}
\maketitle
\end{titlepage}
\baselineskip=21.5pt
\parskip=3pt
\section{Introduction}
One of the distinctive properties of the non-linear $\sigma$-model~\cite{gell},
is that its dynamical variables belong to a non-linear manifold~\cite{zinn},
thus realizing the symmetry group in a non-linear
fashion~\cite{gasi}.  Whence, either the Lagrangian becomes non-polynomial
in terms of unconstrained variables, or it becomes polynomial but in
variables which satisfy a non-linear constraint. It is often convenient
to work in a polynomial or `linearized' representation of the model.
By this we mean an equivalent description where the symmetry is
linearly realized, although the transformations now act on a different
representation space.
There is usually more than one way to construct such linearized
representations. For example, in the $O(N)$ models, where the field
is an $N$-component vector constrained to have constant modulus, a
polynomial representation is constructed simply by introducing a Lagrange
multiplier for that quadratic constraint. However, this simplicity is
not present in general because the constraints required to define the
manifold can be much more complex, like in the $SU(N)$ groups.

In references~\cite{town,cart} a polynomial representation of
the non-linear $\sigma$-model was introduced; let us briefly explain
it for the $SU(N)$ model in d+1 dimensions.

The usual presentation~\cite{slav} of this model is in terms of an
$SU(N)$ field $U(x)$, with Lagrangian density
\begin{equation}
{\cal L} \,=\, \frac{1}{2} g^{d-1} tr( \partial_{\mu}U^{\dagger}
\partial^{\mu}U)\;\;,
\label{01}
\end{equation}
where $g$ is a coupling constant with dimensions of mass (the constant
$f_{\pi}$ in its application to Chiral Perturbation Theory in 3+1
dimensions). The polynomial description~\cite{cart} of this model
is constructed in terms of a non-Abelian ($SU(N)$) vector field $L_{\mu}$
plus a non-Abelian antisymmetric tensor field $\theta_{\mu \nu}$\footnote{
To avoid the proliferation of indices, we frequently work in terms of the
{\em dual} of $\theta_{\mu \nu}$, which in $1+1$ is a pseudo-scalar, in $2+1$
a pseudovector, etcetera.}
with the Lagrangian density
\begin{equation}
{\cal L} \,=\, \frac{1}{2} g^2 L_{\mu} \cdot L^{\mu} +
g \, \theta_{\mu \nu} \cdot F^{\mu \nu}(L)
\label{02}
\end{equation}
where the fields $L_{\mu}$ and $\theta_{\mu \nu}$ are defined by their
components in the
basis of generators of the adjoint representation of the Lie algebra of
$SU(N)$; i.e., $L_{\mu}(x)$ is a vector with components $ L_{\mu}^a, a =
1,\cdots , N^2 - 1$, and analogously for $\theta_{\mu \nu}$. The
components of
$F_{\mu \nu}$ in the same basis are: $F_{\mu \nu}^a (L) = \partial_{\mu}
L_{\nu}^a - \partial_{\nu} L_{\mu}^a + g^{\frac{3-d}{2}} f^{a b c}
L_{\mu}^b L_{\nu}^c $. The dots mean $SU(N)$ scalar product,
for example:
$L_{\mu} \cdot L^{\mu} = \sum_{a = 1}^{N^2 -1} L_{\mu}^a L^{\mu}_a$.
The $d$-dependent exponents in the factors of $g$ are chosen in order to
make the fields have the appropriate canonical dimension for each $d$.

The Lagrange multiplier $\theta_{\mu \nu}$ imposes the constraint
$F_{\mu \nu}(L) =0$, which is equivalent~\cite{itzy} to
$L_{\mu} = g^{\frac{d -3}{2}} U\partial_{\mu}
U^{\dag}$, where $U$ is an element of $SU(N)$. When this is substituted back
in (\ref{02}), (\ref{01}) is
obtained \footnote{For a complete derivation of the
equivalence between the theories defined by (\ref{01}) and (\ref{02})
within the path integral
framework, see ref.~\cite{cart}.}. This polynomial formulation could
be thought of as a concrete Lagrangian realization of the Sugawara theory of
currents~\cite{suga}, where all the dynamics is defined by the currents, the
energy-momentum tensor, and their algebra. Indeed, $L_{\mu}$ corresponds to
one of the conserved currents of the non-polynomial formulation, due
to the invariance of ${\cal L}$ under global (left) $SU(N)$ transformations
of $U(x)$. The energy-momentum tensor for (\ref{02}) is
indeed a function of $L_{\mu}$ only:
\begin{equation}
T^{\mu \nu} \;=\; g^2 ( L^{\mu} \cdot L^{\nu} -
\frac{1}{2} g^{\mu \nu} L^{2} ) \;.
\end{equation}

One  can easily relate amplitudes with external legs of the field
$L_{\mu}$ to the corresponding pions' scattering matrix elements,
as shown in ref.~\cite{slav}.
It is also possible to relate off-shell Green's functions of the field
$U$ to the ones of the field $L_{\mu}$, although this relation
is non-local.  As
\begin{equation}
L_{\mu} \,=\, g^{\frac{d-3}{2}} U\partial_{\mu} U^{\dag} \Rightarrow
D_{\mu}U =0, \;\;
D_{\mu} \,\equiv \, \partial_{\mu} + g^{\frac{3-d}{2}} L_{\mu} \;,
\label{03}
\end{equation}
then $U$ can be obtained at the point $x$ by parallel transporting
its value at spatial infinity, which we fix to be equal to the unit
matrix\footnote{We identify (as usual) all the points at spatial infinity.}:
\begin{equation}
U(x)\,=\, {\cal P} \exp [- g^{ \frac{3-d}{2} }\int_{C_x} dy^{\mu} L_{\mu} (y) ],
\label{04}
\end{equation}
where ${\cal P}$ is the path-ordering operator~\cite{itzy}, and the
line-integral in the exponent is along a curve $C_x$, a regular path
starting at spatial infinity, and ending at $x$. The condition
$F_{\mu\nu}=0$
guarantees that $U$ is in fact invariant under deformations of $C_x$
which leave its endpoints unchanged.
We can also construct products of two or more fields
in a similar way, for example
\begin{equation}
U(x_2) \, U^{-1} (x_1) \;=\;{\cal P} \exp [ -
g^{ \frac{3-d}{2} } \int_{C_{x_1 \to x_2}}
dy_{\mu}
L^{\mu} (y) ] \;,
\label{05}
\end{equation}
where $C_{x \to y}$ is a continuous path from $x_1$ to $x_2$. This
shows how $U$-field correlation functions can in principle be
calculated using Lagrangian (\ref{02});
one has to evaluate, for example, the Wilson line (\ref{05}) in
the theory defined by (\ref{02}).

The classical equations of motion for the Lagrangian (\ref{02})
are
\begin{eqnarray}
L^{\nu} (x) &=& \frac{1}{g} D_{\mu} \theta^{\mu \nu}(x) \;,\nonumber\\
F^{\mu \nu} (L) &=& 0 \;.
\label{06}
\end{eqnarray}
Taking the covariant divergence on both sides of the first equation
of motion, and using the second one, one gets
\begin{equation}
\partial \cdot L (x) \,=\, 0 \;.
\label{07}
\end{equation}
Inserting $L_{\mu} = g^{\frac{3-d}{2}} U\partial_{\mu}
U^{\dagger}$  in (\ref{07}), it yields the equations of motion for the
usual non-polynomial Lagrangian (\ref{01}). Note that the solutions
of (\ref{06}) will, in general, contain arbitrary functions of the
time. If we know a solution, performing on it the transformation:
\begin{eqnarray}
\theta_{\mu \nu}(x) &\to& \theta_{\mu \nu} (x) + \delta_{\omega}
\theta_{\mu \nu} (x) \nonumber\\
\delta_{\omega} \theta_{\mu \nu} (x) &=& D^{\rho} \omega_{\rho \mu \nu}(x)
\;,
\label{08}
\end{eqnarray}
where $\omega_{\rho \mu \nu}(x)$ is an arbitrary completely antisymmetric
tensor field, will produce another solution, because $D_{\mu}
\delta_{\omega} \theta^{\mu \nu}$ vanishes as a consequence of the
Bianchi identity for $L_{\mu}$\footnote{This kind of symmetry also
appears when considering the dynamics of a two-form gauge field,
see for example references~\cite{anti}}.
Obviously $d$ must be larger than one in order to this transformation be
well defined, since at least three different indices are needed
to have a Bianchi identity.
This degeneracy in the equations of motion is due to the
gauge invariance of the action under the transformations (\ref{08}).

This gauge-invariance makes the quantization of the model
interesting, and it will allow us to discuss some properties
of the non-linear $\sigma$-model from the (unusual) point of view of gauge
systems.
The Hamiltonian formulation of the model possesses a rich structure,
since there are second-class constraints (${\cal L}$ is first-order),
first-class constraints (for $d > 1$), and moreover they are
reducible for $d > 2$.

The structure of the paper is as follows: In section 2 we
discuss the Hamiltonian formulation of the $1+1$, $2+1$ and $3+1$
models, following the Dirac algorithm~\cite{dira}. In section 3 we construct the
general gauge invariant functionals for the transformations generated
by the first-class constraints found in section 2, and in
section 4 we apply the Dirac's brackets method to the second-class
system formed by the first-class constraints plus some
canonical gauge-fixing conditions. In section 5 we present our
conclusions.

In Appendix we discuss a duality relationship
between first-class systems, which generalizes a property we discuss
for the $2+1$-dimensional model.

\section{Hamiltonian formalism and constraints}

\subsection{$1+1$ dimensions}

From Section 1, the polynomial Lagrangian in $1+1$ dimensions becomes
\begin{equation}
{\cal L} \;=\; \frac{1}{2} g^2 L_{\mu} L^{\mu} + \frac{1}{2}
\, g \, \theta \, \epsilon_{\mu \nu} F^{\mu \nu} (L)
\label{11}
\end{equation}
where $\theta$ is a pseudoscalar field. It is evident
that there is no gauge symmetry in this case.
Thus there will not be first-class constraints in the
Hamiltonian formulation. However, there
are second-class constraints, because ${\cal L}$ is of
first-order in the derivatives. This property will also appear
in higher dimensions, so we will only discuss it
in some detail for this case.
To start with, we rewrite (\ref{11}) in a more explicit form
\begin{equation}
{\cal L} \,=\, \frac{1}{2} g^2 L_0^a L_0^a - \frac{1}{2} g^2 L_1^a
L_1^a + g \theta^a \partial_0 L_1^a - g \theta^a \partial_1 L_0^a +
g^2 \theta^a f^{a b c} L_0^b L_1^c \;.
\label{12}
\end{equation}
Next we define the canonical momenta, where the
primary constraints appear:
\begin{eqnarray}
{\pi}_0^a (x) &\equiv& \frac{ \partial {\cal L} }{\partial (\partial_0 L_0^a)}
\approx 0
\nonumber\\
{\pi}_1^a &\equiv& \frac{ \partial {\cal L} }{\partial (\partial_0 L_1^a)}
\approx g \theta^a (x)
\nonumber\\
\pi^a_{\theta} (x) &\equiv& \frac{ \partial {\cal L} }{\partial (
\partial_0 \theta^a)}
\approx 0
\label{15}
\end{eqnarray}
and the canonical Hamiltonian becomes
\begin{equation}
H \,=\, \int d {\bf x} [ - \frac{1}{2} g^2 L_0^a L_0^a + \frac{1}{2} g^2
L_1^a L_1^a - g L_0^a {(D_1 \theta )}^a ] \;,
\label{16}
\end{equation}
where
\begin{equation}
{(D_1 \theta)}^a \,=\, \partial_1 \theta^a + g f^{a b c} L_1^b \theta^c
\;.
\label{17}
\end{equation}

The `total' Hamiltonian is constructed as usual, adding to (\ref{16})
a Lagrange multiplier term for each of the primary constraints
(\ref{15}). Following the Dirac's algorithm one more
constraint is obtained:
\begin{equation}
g L_0^a (x) \approx - {( D_1 \theta)}^a (x)
\label{18}
\end{equation}
and the Lagrange multipliers become fully determined. The full set of
(primary plus secondary) constraints is second-class, and its particular
form allows us to eliminate the canonical pairs of
$L_0^a$ and $\theta^a$, thus effectively eliminating the associated degrees
of freedom. The Dirac bracket becomes equal to the Poisson bracket
for the remaining degrees of freedom. The resulting Hamiltonian is
\begin{equation}
H \;=\; \int d x [ \frac{1}{2 g^2} D_1 \pi_1 \cdot D_1 \pi_1 \,+
\frac{1}{2} g^2 L_1 \cdot L_1 ]\;,
\label{19}
\end{equation}
with canonical brackets between the $L_1^a$'s and their momenta $\pi_1^a$.
Thus these two variables become symplectic coordinates on the reduced
phase-space, or constraint surface.

\subsection{$2+1$ dimensions}
The polynomial Lagrangian in this case becomes,
\begin{equation}
{\cal L} \;=\; \frac{1}{2} g^2 L_{\mu} \cdot L^{\mu} + \frac{1}{2}
\, g \, \theta_{\mu} \,\cdot \epsilon^{\mu \nu \lambda} F_{\nu \lambda} (L).
\label{121}
\end{equation}
The constraint algorithm\footnote{The full details of the application
of the Dirac algorithm to this system will be presented elsewhere.} produces
the $2+1$ analogous of the
second-class constraints we showed in Section 1, allowing us to eliminate
the $0$-component of $L_{\mu}$ and all the components of
$\theta_{\mu}$. However, there will remain a set of first-class
constraints
\begin{equation}
G^a (x) \;=\; \frac{1}{2}  \epsilon_{j k} F_{j k}^a \; \approx \; 0\;,
\label{122}
\end{equation}
with the first-class Hamiltonian
\begin{equation}
H \;=\; \int d^2 x [ \frac{1}{2 g^2} D_j \pi_j \cdot D_k \pi_k \, + \,
\frac{1}{2} g^2 L_j \cdot L_j ]\;.
\label{123}
\end{equation}
They satisfy the algebra
\begin{eqnarray}
\{ G^a (x) , G^b (y) \} &=& 0 \nonumber\\
\{ H , G^a (x) \} &=& V^{a b} G^b (x) \nonumber\\
V^{a b} &\equiv& g^{- \frac{3}{2} } f^{a c b} {(D_j \pi_j)}^c (x).
\label{124}
\end{eqnarray}

Now we show in what sense we can relate the massless Yang-Mills
theory to the non-linear $\sigma$-model in this formulation. The
$SU(N)$ Yang-Mills theory is defined by the Lagrangian
\begin{equation}
{\cal L}_{YM} \;=\; - \frac{1}{4} F_{\mu \nu}(L) \cdot F^{\mu \nu}
(L) \;,
\label{125}
\end{equation}
which in the temporal gauge gives rise to the canonical Hamiltonian
\begin{equation}
H \;=\; \int d^2 x [ \frac{1}{2} \pi_j \cdot \pi_j + \frac{1}{4}
F_{j k} \cdot F_{j k} ] \;,
\label{126}
\end{equation}
and the first-class constraints (`Gauss' laws'):
\begin{equation}
H_a(x) \;=\; (D_j \pi_j )_a(x) \approx 0 \;,
\label{127}
\end{equation}
which satisfy the $SU(N)$ algebra
\begin{equation}
\{ H_a (x) , H_b (y) \} \;=\; \delta (x-y) f_{a b c} H_c (x) \;.
\label{128}
\end{equation}
Note that (\ref{126}) can be rewritten as
\begin{equation}
H \;=\; \int d^2 x [ \frac{1}{2} \pi_j \cdot \pi_j + \frac{1}{2}
G \cdot G ] \;,
\label{129}
\end{equation}
where the $G_a$'s are the ones defined in (\ref{122}).
We then see that the first-class systems corresponding to the Yang-Mills
model and the non-linear $\sigma$-model can be related by: 1) Interchanging
the constraints:
\begin{equation}
H_a (x) \leftrightarrow G_a (x)\;,
\label{130}
\end{equation}
and 2) Interchanging  $L_j$ by $\frac{1}{g} \pi_j$ in the non-derivative
terms in the Hamiltonian.
The generalization of this mapping is constructed in Appendix.

\subsection{$3+1$ dimensions}
After eliminating the second-class constraints, one obtains a first-class
Hamiltonian which looks exactly like the one of the
$2+1$-dimensional case:
\begin{equation}
H \;=\; \int d^3 x [ \frac{1}{2 g^2} D_j \pi_j \cdot D_k \pi_k \, + \,
\frac{1}{2} g^2 L_j \cdot L_j ]\;,
\label{131}
\end{equation}
and the set of first-class constraints
\begin{equation}
G_j^a (x) \;=\; \frac{1}{2}\epsilon_{j k l} F_{k l}^a (x) \approx 0
\;.
\label{132}
\end{equation}
Although the system seems to be the obvious generalization of the
$2+1$-dimensional one,
there is an essential difference: The constraints (\ref{132})
are not all independent, but verify the Bianchi identity:
\begin{equation}
(D_j G_j)^a (x) \;=\; 0 \;\;,\;\; \forall a \;.
\label{133}
\end{equation}
This implies that the set of constraints is reducible, containing
only two independent functions. The counting of degrees of freedom
then gives $1$ for the number of physical dynamical variables ($1 =
3 - 2$).
We mention that the elimination of the second-class constraints applies
in a similar way to the general $d$-dimensional case, and that the
Hamiltonian and constraints are the obvious generalizations
of (\ref{131}) and (\ref{132}), respectively. Due to the existence
of the Bianchi identity in general, the number of independent
constraints in an arbitrary dimension is just enough to kill $d-1$ out
of the $d$ degrees of freedom in $H$, leading to only one physical
variable, as it should be for a model which describes the dynamics
of a scalar field.

\section{Gauge invariant functionals}
Gauge invariant functionals\footnote{We assume the denomination
`gauge-invariant' to mean {\em on-shell} gauge invariance, i.e.,
the gauge-invariant functionals are invariant on the constraint
surface.} are important
from both the classical
and quantum mechanical points of view. Classically, a complete set of
gauge invariant functionals and their equations of motion completely
determines the dynamics of the observable, i.e., {\em physical} degrees
of freedom. In Quantum Mechanics, Dirac's method for first-class constraints
defines the `physical' subspace of the complete  Hilbert space as
the one whose state vectors are annihilated by the first-class
constraints, i.e., the gauge-invariant ones. In the Schroedinger
representation, the physical subspace consists of gauge
invariant functionals of the fields.

To construct the gauge invariant functionals, we make use of the
concept of gauge invariant projection, defined as follows:
Let $I\,=\,I[\pi,L]$ be an arbitrary functional of the phase-space
fields. Then its gauge-invariant projection ${\cal P}(I)[\pi,L]$ is defined by:
\begin{equation}
{\cal P}(I) [\pi,L] \;=\; \frac{1}{{\cal N}}
\int {\cal D} \omega \, I [ \pi^{\omega} , L ]\;,
\label{21}
\end{equation}
where $\pi^{\omega}$ is the gauge-transformed of $\pi$ by the gauge
group element $\omega (x)$ (for example, in  $2+1$ dimensions,
$\pi_j^{\omega} = \pi_j + \epsilon_{j k} D_k \omega$) and
the functional integration is over all the possible configurations
for $\omega$. The normalization factor ${\cal N}$ is just the
volume of the gauge group: ${\cal N} \,=\, \int {\cal D} \omega$.
It is then easy to see that the gauge invariant projection of
an arbitrary functional is indeed gauge invariant:
\begin{equation}
{\cal P}(I) [ \pi^{\omega} , L] \;=\; {\cal P}(I) [ \pi , L] \;,
\label{22}
\end{equation}
and that ${\cal P}$ is a linear projection operator:
\begin{eqnarray}
{\cal P} (\lambda_1 I_1 + \lambda_2 I_2 )  &=& \lambda_1 {\cal P} (I_1) +
\lambda_2 {\cal P}(I_2) \;,
\nonumber\\
{\cal P}^2 &=& {\cal P} \;\;,\;\; \forall I \;.
\label{23}
\end{eqnarray}
A functional $F$ is gauge invariant if and only if
${\cal P}(F) = F$. This can be shown to be equivalent to saying
that $F$ belongs to
the image of ${\cal P}$. We then construct the most general
gauge-invariant functional by applying ${\cal P}$ to an arbitrary
functional.

In $2+1$ dimensions, we further decompose the momentum as
\begin{equation}
\pi_j (x) \;=\; D_j \alpha (x) + \epsilon_{j k} D_k \beta (x) \;,
\label{24}
\end{equation}
(where $\alpha$ and $\beta$ are scalar and pseudoscalar, respectively)
to show that
\begin{eqnarray}
{\cal P}(I)[\pi,L] &=& {\cal P}(I) [\alpha, \beta, L] \;=\;
\frac{1}{{\cal N}} \int {\cal D} \omega \; I[D_j \alpha + \epsilon_{j k}
D_k ( \beta + \omega) , L] \nonumber\\
&=& I[ \alpha, 0, L]
\label{25}
\end{eqnarray}
where the last line was obtained by performing the shift
$\omega \to \omega - \beta$. (\ref{25}) shows that any gauge invariant
functional is independent of $\beta$; the reciprocal is immediate.
The conclusion can be put as follows: The general gauge invariant
functional depends arbitrarily on $L$, an on $\pi$ only through the
combination $D_j \pi_j$.

This result is generalizable to $3+1$ dimensions. $F$ is shown to
depend only on $D_j \pi_j$ and $L_j$, by using the same argument as
in the $2+1$ case. The decomposition of $\pi_j$ is now
\begin{equation}
\pi_j (x) \;=\; D_j \alpha (x) + \epsilon_{j k l} D_k \beta_l (x)
\;,
\label{26}
\end{equation}
and the $\beta$ dependence is removed as before by a shift in $\omega$.
The only difference appears in the actual construction of the projection
operator, which appears to be ill-defined at first sight. This is so
because the gauge transformations in $d = 3$:
\begin{equation}
\pi^{\omega}_j \;=\; \pi_j + \epsilon_{j k l} \, D_k \, \omega_l \;,
\label{27}
\end{equation}
are invariant under $\omega_j(x) \to \omega_j(x) + D_j \lambda (x)$,
for any $\lambda$.
This produces an infinite factor when one integrates over $\omega$ in
the definition (\ref{21}) of ${\cal P}(I)$. Of course, this factor is
also present
in ${\cal N}$, but to explicitly cancel them on needs to `fix the gauge'
for the integration over $\omega$. A convenient way to do that is by
using the Faddeev-Popov trick, which gives the
`gauge fixed' projector
\begin{eqnarray}
{\cal P}(I)[\pi,L] &=& \frac{1}{\int  {\cal D} \omega \det M_f [\omega]
\delta [f(\omega)] } \nonumber\\
&\times& \int {\cal D} \omega \det M_f [\omega] \delta [f(\omega)]
I[ D_j \alpha + \epsilon_{j k l} D_k \omega_l ] \;,
\label{28}
\end{eqnarray}
where $M_f [\omega]= \frac{\delta}{\delta \lambda}f(\omega^{\lambda})$.
We have seen that the gauge invariant functionals depend on $D_j \pi_j$
and $L_j$ (the result is indeed true in any number of dimensions).
However, there is still a degree of redundancy in this description
because one is interested only in gauge invariant functions {\em
on-shell}, i.e., on the surface $F_{j k} (L) = 0$. Thus
we do not need the full $L_j$, but only its restriction
to the constraint surface.
As it was shown in ref.~\cite{slav}, it is possible to solve that
kind of equation using a perturbative approach. The main result
we need to recall is that that perturbative expansion allows one
to express $L_j$ as a function of the scalar $\partial_j L_j$ only.
Then we obtain a more symmetrical description in terms of the
gauge-invariant, scalar variables:
\begin{equation}
(D_j \pi_j)^a (x) \;\;,\;\; (D_j L_j)^a (x) = \partial_j L_j^a (x)\;.
\label{29}
\end{equation}
Their equations of motion link each other:
\begin{eqnarray}
\frac{\partial}{\partial t} (D_j \pi_j) &=& - g^2 \partial_j L_j
\nonumber\\
\frac{\partial}{\partial t} (\partial_j L_j) &=& - g^{-2} \partial_j D_j
(D_k \pi_k) \nonumber\\
F_{j k} (L) &\approx& 0\;,
\end{eqnarray}
(where we have included the constraints).
They imply the second order equations
\begin{eqnarray}
( \partial_t^2 - \partial_j D_j ) D_k \pi_k &=& 0 \nonumber\\
( \partial_t^2 - \partial_j D_j ) \partial_k L_k &=& 0,
\label{30}
\end{eqnarray}
which show the scalar particle nature of the (only) physical degree of freedom.
Let us consider in more detail the issue of {\em static} solutions in
$3+1$ dimensions. In this situation, (39) reduces to
\begin{eqnarray}
\partial_j L_j &=& 0 \;\;\;, \;\;\; F_{j k} \;\approx\; 0 \nonumber\\
\partial_j D_j ( D_k \pi_k ) &=& 0 \;.
\label{31}
\end{eqnarray}
The first two equations in (\ref{31}) are equivalent to
\begin{equation}
L_j \;=\; U\partial_j U^{\dagger} \;\; , \;\; \partial_j L_j
\,=\,0
\;.
\label{32}
\end{equation}
They are exactly the set of equations one gets when considering the
Gribov problem~\cite{grib}(for the Yang-Mills theory) in the Coulomb gauge,
on the orbit of
the trivial configuration ($L_j =0$). It is well known that there
are more solutions than just the trivial one, in particular, one
obtains the `fermionic' configurations of the Skyrme model\footnote{The
stabilizing term can be added without changing the
canonical structure of the model.},
which verify
\begin{equation}
n \;=\; - \frac{1}{24 \pi^2} \int d^3 x \;\epsilon_{j k l} \;
tr ( L_j L_k L_l)\;=\; \pm \frac{1}{2} \;.
\label{33}
\end{equation}
Once a particular solution of (\ref{33}) is obtained,
it can be inserted in the last equation of (\ref{31})
to get an equation for $\pi$. Note that the momenta should then
satisfy
\begin{equation}
D_j \pi_j \;=\; f_0  \;,
\label{34}
\end{equation}
where $f_0$ is a zero mode of the operator $\partial_j D_j$ (of
course, the trivial solution $\pi_j = 0$ is included). For each
Gribov solution $L_j$, there will be a non-trivial zero mode
for this operator, and then a non-zero solution for the momenta. These
solutions can be compared to the static solutions of the usual
non-polynomial formulation. To do that we must regain the field
$L_0$, which was eliminated by using the second-class constraints.
That is very simple, since in fact $L_0$ is equal to a constant
times $D_j \pi_j $, and then (\ref{34}) implies
\begin{equation}
L_0 \; =\; f_0 \;.
\label{35}
\end{equation}
So the family of static solutions in the polynomial version seems
to be larger than in the usual treatment.  Indeed, as $L_0 =
U \partial_0 U^{\dagger}$, a non-zero $L_0$ implies that there is
a time-dependence for $U$. Note, however, that such configurations
contribute to the energy in an amount:
\begin{equation}
E(L_0) \;=\; \frac{1}{2} g^2  \int d^3 x {[f_0 ]}^2 \;,
\label{36}
\end{equation}
which is proportional to the norm of the zero mode, and then
the minimum energy will correspond to the trivial
configuration $L_0 =0$.
A simple example of a configuration with $L_0 \neq 0$ is:
\begin{eqnarray}
{\tilde L}_j ({\bf x} , t) &=& \exp (i h t) L_j({\bf x}) \exp (-i h t)
\nonumber\\
{\tilde L}_0 (t) &=& h \;,
\label{37}
\end{eqnarray}
where $h$ is a hermitian (constant) traceless matrix, and
$L_j({\bf x})$ satisfies (\ref{32}). Thus for
(\ref{37}), $E(L_0)\,=\,\frac{1}{2} g^2 tr(h^2) \int d^3 x$, which
is divergent for infinite volume.

\section{Dirac's brackets method}
As an alternative to the previous approach, we apply here
the `Dirac's brackets method' to the treatment of the first-class
constraints in the $2+1$ model (it can however be straightforwardly
generalized to the $d+1$ model). It consists in constructing the Dirac's
brackets for the set of
{\em second-class} constraints containing all the original first-class
constraints plus a suitable set of gauge fixing conditions.
We choose the canonical gauge fixing functions:
\begin{equation}
\chi^a (x) \;=\;  \pi_2^a (x) \;=\; 0 \;.
\label{ss1}
\end{equation}
The basic ingredient to calculate the Dirac's brackets is the
Poisson bracket between $\chi^a$ and $G^a (x)$:
$\{ \chi^a (x) , G^b (y) \} = (D_1)^{a b} \delta ({\bf x} - {\bf y}) $.
From this it follows that the only non-trivial Dirac's brackets
between canonical variables are
\begin{eqnarray}
\{ L_1^a (x) , \pi_1^b (y) \}_D &=& \delta_{a b} \delta ({\bf x} -{\bf y})
\;,\nonumber\\
\{ L_2^a (x) , \pi_1^b (y) \}_D &=& \langle x,a\mid
D_1^{-1} D_2 \mid y,b \rangle \;.
\label{ss2}
\end{eqnarray}
The second one is a complicated non-local function. It is more
convenient to take advantage of the results of the previous section to
work with $L_j$ and $D_1 \pi_1$. Then the Dirac's brackets become
local
\begin{eqnarray}
\{ L_1^a (x) , (D_1 \pi_1)^b (y) \}_D &=& - D_1^{a b} \delta ({\bf x}-
{\bf y})
\nonumber\\
\{ L_2^a (x) , (D_1 \pi_1)^b (y) \}_D &=& - D_2^{a b} \delta ({\bf x} -
{\bf y} ) \;.
\label{ss3}
\end{eqnarray}

\section{Conclusions}
The polynomial formulation (\ref{02}) has an interesting canonical
structure. Some of its properties are:
\begin{description}
\item The system has second-class constraints which can be solved
explicitly for some coordinates in terms of the others. This leaves the
canonical pairs associated to the spatial components of a
non-Abelian vector field only.
\item For $d>1$  there remain first-class constraints which
form an Abelian algebra. They, and
the first-class Hamiltonian have essentially the same structure
in any number of dimensions.
However, for $d>2$, the constraints are reducible. The number of
independent constraints is just enough to leave only one physical
degree of freedom.
\item These first-class systems can be regarded as `duals' of the
Yang-Mills model in the temporal gauge, in the sense that
the constraints in one of the systems are non-trivial gauge
invariant functions in the other. This  duality can be generalized
to a greater class of first-class systems.
\item We also constructed the most general gauge invariant
functional explicitly. Note that the Gauss-law constraints
of the dual Yang-Mills system appear here as (non-trivial) gauge
invariant objects, verifying the general property discussed in
the Appendix.
\end{description}

\section*{Acknowledgements}
C. D. F. was supported by an European Community Postdoctoral Fellowship.
T. M. was supported in part by the Daiwa
Anglo-Japanese Foundation. We also would like to express our
acknowledgement to Dr. I. J. R. Aitchison for his kind hospitality.

\newpage

\appendix
\section*{Appendix: A duality transformation for first-class systems}
The kind of `duality' that exists between the massless Yang-Mills
theory and the polynomial version of the non-linear $\sigma$-model
is a particular case of a more general concept, which we define in
this Appendix. Let us consider a constrained dynamical system defined on a
phase-space of coordinates $q_j, p_j, \,\,\, j=1 \cdots N$, with first-class
Hamiltonian $H$ and a complete set of irreducible first-class constraints
$G_a \approx 0 , \,\,\, a=1, \cdots ,N$. We assume that the first-class
constraints satisfy the closed algebra
\begin{equation}
\{ G_a , G_b \} \;=\; g_{a b c} (q,p) \, G_c
\label{a1}
\end{equation}
and regarding the Hamiltonian, we impose on it the requirement of
having the structure
\begin{equation}
H \; = \; \frac{1}{2}\,\, F_a (q,p) \, F_a (q,p)
\label{a2}
\end{equation}
where the functions $F_a (q,p),\,\,\, a=1, \cdots ,N$ verify the relations
\begin{eqnarray}
\{ F_a , F_b \} &=& f_{a b c} (q , p) \,\, F_c \nonumber\\
\{ G_a , F_b \} &=& {\lambda}_{a b c} (q , p) \,\, F_c
\label{a3}
\end{eqnarray}
and $\lambda$ is completely antisymmetric with respect to the
last two indices. This implies that the Poisson bracket of $H$ and
each of the $G_a$'s will be {\em strongly} equal to zero, what is
stronger than what we need in a general first-class system.
Indeed, Equations (\ref{a2}) and (\ref{a3}) select among all the
possible first-class systems the class which admit a duality of the
kind we are going to define.

The associated dual first-class system is defined on the same phase-space,
and its Hamiltonian and constraints (denoted with a tilde) are defined by
\begin{eqnarray}
{\tilde H} \,&=&\, \frac{1}{2} \,\, G_a \, G_a \nonumber\\
{\tilde G}_a \,&=&\, F_a \, \approx \, 0 \;,
\label{a4}
\end{eqnarray}
where $F_a$ and $G_a$ are the ones introduced in (\ref{a1}), (\ref{a2})
and (\ref{a3}). We then verify that the new system is also first-class, since
\begin{eqnarray}
\{ {\tilde G}_a , {\tilde G}_b \} &=& {\tilde g}_{a b c} (q , p) \,\,
{\tilde G}_c \nonumber\\
\{ {\tilde G}_a , {\tilde H} \} &=& V_{a b} (q , p) \,\, {\tilde G}_b
\;,
\label{a5}
\end{eqnarray}
where:
\begin{eqnarray}
{\tilde g}_{a b c } (q,p) &=& f_{a b c} (q,p) \nonumber\\
V_{a b} (q,p)  &=& \lambda_{a b c}(q,p) \,\, G_c (q,p) \;.
\label{a6}
\end{eqnarray}
Thus evidently this mapping leaves the first-class nature of
the system invariant. Note however, that the irreducibility of the
new constraints is by no means guaranteed. That will depend upon the
particular form of the $F_a$'s.
An interesting property of the new system is that, because of (\ref{a3}),
\begin{equation}
\{ G_a , {\tilde G_b} \} \; \approx \; 0 \;\; \forall a, b \; ,
\label{a7}
\end{equation}
which proves that the $G_a$'s constitute a set of $M$ independent
gauge invariant functions, which is a very helpful property when one
wants to study the classical or quantal dynamics of the system.

Note that the transformation we defined is not necessarily involutive;
to guarantee that we would need a completely antisymmetric $\lambda_{a b
c}$ in Equation (\ref{a3}).

The Hamiltonian (\ref{a2}) resembles the one of the Yang-Mills
system, except for the absence of the term quadratic in the
canonical momenta. We did not include this, neither the
corresponding one
in the dual, to keep the discussion as general as possible. As
they are gauge invariant by themselves,
their presence or not do not alter the essence of the discussion.

This duality transformation can be interpreted as
transforming a gauge-invariant theory into another. The unphysical gauge
variables of the first theory come to be the true physical degrees of
freedom of the second one, which are actually defined on the fibers
generated by the gauge group of the first model (note that the second
theory is not invariant under the gauge group of the first one).  Hence it
follows naturally that the first-class constraints which generate the
gauge transformations of the first theory are the dynamical variables
which describe  motions (`translations') along the gauge group fiber.

This duality transformation may be useful  in cases
where the number of physical degrees of freedom is the same in both the
original and dual models, as for the Yang-Mills and non-linear sigma-model
in 2+1 dimensions. Here the physical excitations are massless scalar
fields for both models. Although it is very difficult to write down
an explicit dynamics for the physical degree of freedom of the
Yang-Mills theory, this mapping could make it easier, since the
identification of physical variables is much simpler in the non-linear
sigma-model.


\newpage
\begin{thebibliography}{100}
\bibitem{gell}For early references, see for example:
M. Gell-Mann and M. L\'evy, Nuovo Cimento {\bf 16},
705 (1960);

S. Weinberg, Phys. Rev. {\bf 166}, 1568 (1968);

J. Schwinger, Phys. Rev. {\bf 167}, 1432 (1968).
\bibitem{zinn} J. Zinn-Justin, {\em Quantum Field Theory and
Critical Phenomena}, Oxford University Press, Second Edition (1993).
\bibitem{gasi}S. Gasiorowicz and D. Geffen,
Rev. Mod. Phys. {\bf 41},
531 (1969);

S. Coleman, J. Wess and B. Zumino, Phys. Rev {\bf 177}, 2239 (1969).
\bibitem{town}D. Z. Freedman and P. K. Townsed, Nucl. Phys.
B{\bf 177}, 282 (1981);

N.K. Nielsen, Nucl. Phys. B{\bf 332}, 391 (1990).
\bibitem{cart}G. L. Demarco, C. D. Fosco and R. C. Trinchero,
Phys. Rev D{\bf 45}, 3701 (1992). See also:
C. D. Fosco and R. C. Trinchero, Phys. Lett. B{\bf 322}, 97 (1994);
C. D. Fosco and T. Matsuyama, Phys. Lett. B{\bf 329}, 233 (1994);
Int. J. Mod. Phys. A{\bf 10}, 1655(1995).
\bibitem{dira}P. A. M. Dirac, Can. J. Math. {\bf 2}, 129 (1950);
ibid. {\bf 3}, 1 (1951); Proc. Roy. Soc. (London), A{\bf 246},
326 (1958); `Lectures on Quantum Mechanics', Yeshiva University,
New York, Academic Press (1967).
\bibitem{suga}H. Sugawara, Phys. Rev. {\bf 120}, 1659 (1968).
\bibitem{slav}A. A. Slavnov, Nucl. Phys. B{\bf 31}, 301 (1971).
\bibitem{itzy}C. Itzykson and J. B. Zuber, {\em Quantum Field Theory},
McGraw-Hill (1986).
\bibitem{flan}H. Flanders, {\em Differential Forms}, Academic Press,
New York (1963).
\bibitem{anti} W. Siegel, Phys. Lett. B{\bf 93}, 170 (1980);
J. Thierry-Mieg, Nucl. Phys B{\bf 335}, 334 (1990).
\bibitem{grib} V. N. Gribov, Nucl. Phys. B{\bf 335}, 334 (1990).
\end{thebibliography}
\end{document}